\documentclass{amsart}[11pt]
\usepackage{amsmath}
\usepackage{amssymb}
\usepackage{amsthm}
\usepackage{subfigure}
\usepackage{mathrsfs}
\usepackage[margin=1in]{geometry}
\usepackage{color}
\usepackage{graphicx}
\usepackage{amsfonts}
\usepackage{xcolor}
\usepackage{hyperref}
\usepackage{enumerate}
\usepackage{xcolor}
\usepackage{tikz}
\usepackage{mathtools }
\usetikzlibrary{decorations.pathreplacing, positioning}

\newtheorem{theorem}{Theorem}[section]

\theoremstyle{definition}

\newtheorem{remark}[theorem]{Remark}

\usepackage{algorithm2e}
\usepackage{multirow}
\usepackage{type1cm}
\usepackage{lmodern}

\newcommand{\fHat}{\widehat{f}}
\newcommand{\fTilde}{\widetilde{f}}
\newcommand{\gHat}{\widehat{g}}
\newcommand{\Mm}{\mathcal{M}}
\newcommand{\Nn}{\mathcal{N}}
\newcommand{\Oo}{\mathcal{O}}
\newcommand{\Pp}{\mathcal{P}}
\newcommand{\ch}{\widehat{c}}
\newcommand{\ab}{\boldsymbol{a}}
\newcommand{\alb}{\boldsymbol{\alpha}}

\newcommand{\chbf}{\widehat{\boldsymbol{c}}}
\newcommand{\cbf}{\boldsymbol{c}}
\newcommand{\Xb}{\boldsymbol{X}}
\newcommand{\xbf}{\boldsymbol{x}}

\newcommand{\uu}{\boldsymbol{u}}

\newcommand{\RR}{\mathbb{R}}
\newcommand{\NN}{\mathbb{N}}

\newcommand{\EE}[1]{\mathbb{E} \left[ #1 \right]}

\newcommand{\BS}{\mathrm{BS}}
\newcommand{\D}{\mathrm{d}}
\newcommand{\Pif}{\boldsymbol{\Pi}}
\newcommand{\Wf}{\boldsymbol{W}}
\newcommand{\Bf}{\boldsymbol{B}}
\newcommand{\CPi}{\boldsymbol{\breve{\Pi}}}
\newcommand{\E}{\mathrm{e}}

\DeclareMathOperator*{\argmin}{arg\,min}

\title{Stacked Monte Carlo for option pricing}

\author{Antoine Jacquier}
\address{Department of Mathematics, Imperial College London, London SW7 1NE, UK}
\email{a.jacquier@imperial.ac.uk}
\author{Emma R. Malone}
\address{Lloyds Banking Group plc, Commercial Banking, 10 Gresham Street, London, EC2V 7AE,~UK}
\email{Emma.Malone@lloydsbanking.com}
\author{Mugad Oumgari}
\address{Lloyds Banking Group plc, Commercial Banking, 10 Gresham Street, London, EC2V 7AE,~UK}
\email{Mugad.Oumgari@lloydsbanking.com}
\date{\today}
\thanks{The views and the opinions expressed here are those of the authors and do not represent the opinions of their employers. 
They are not responsible for any use that may be made of these contents. 
No part of this paper is intended to influence investment decisions or promote any product or service.
The authors are grateful to Romain Palfray for lending his expertise on option pricing and for the many helpful suggestions. 
We thank Manlio Trovato, Lee McGinty, Arjun Nayyar, and Helmut Glemser for reviewing and supporting this work.
}

\begin{document}
\maketitle

\begin{abstract}
We introduce a stacking version of the Monte Carlo algorithm in the context of option pricing.
Introduced recently for aeronautic computations, this simple technique,
in the spirit of current machine learning ideas, learns control variates 
by approximating Monte Carlo draws with some specified function. 
We describe the method from first principles and suggest appropriate fits,
and show its efficiency to evaluate European and Asian Call options in constant and stochastic volatility models. 
\end{abstract}

%%%%%%%%%%%%%%%%%%%%%%%%%%%%%%%%%%%%%%%%%%%%%%
\section{Introduction}
Monte Carlo methods are the most fundamental pillar for pricing models in quantitative finance,
and a considerable amount effort has been made to refine their properties, 
from variance reduction techniques~\cite[Chapter 4]{Glasserman} 
(importance sampling, antithetic variables or control variates)
to discretisation of stochastic differential equations~\cite{Kloeden}
and their multilevel extensions~\cite{Giles1, Giles2, Giles3}.
Indeed, most financial instruments do not admit closed-form pricing formulae,
and any numerical technique is thus at the mercy of instabilities and approximation errors. 
This of course generated an appetite for variance reduction techniques, 
whereby more stability can be achieved with similar computation time and level of accuracy. 
There has recently also been a lot of interest in leveraging the power of machine learning tools, 
mainly on the use of neural networks to solve pricing~\cite{Madan, Ferguson, Ghee, Hahn, Hutchinson, Kondratyev} 
and calibration~\cite{Bayer, Horvath, Stone} problems. 

We consider here an alternative approach, which involves the application of simple regression techniques to improve Monte Carlo estimates. We borrow the idea from Alonso, Tracey and Wolpert~\cite{Tracey, Tracey2} 
who were first motivated by applications in aeronautics. 
The main idea is to learn appropriate control variates for option pricing problems, 
in order to achieve high levels of variance reduction.
This task faces two challenges; the first is to obtain unbiased estimates of the pricing payoff function, 
as the Stacked Monte Carlo technique proposed in~\cite{Tracey,Tracey2} did not apply to derivatives pricing. 
The second challenge is to select the appropriate fitting functions for our pricing problems, 
such that the model closely approximates the solution whilst being sufficiently simple to minimise computation times.
We start below by presenting the stacked Monte Carlo in its simple form, namely to numerically compute 
integrals, before applying it pricing options in the Black-Scholes and in the Heston model.
We shall consider European options as well as Asian options, showing how to adapt the method to 
multivariate problems.
We then carry out numerical tests highlighting both the simplicity and the efficiency of the method.

%%%%%%%%%%%%%%%%%%%%%%%%%%%%%%%%%%%%%%%%%%%%
%%%%%%%%%%%%%%%%%%%%%%%%%%%%%%%%%%%%%%%%%%%%
\section{Stacked Monte Carlo}
The Stacked Monte Carlo (StackMC) method~\cite{Tracey, Tracey2}, is a post-processing technique for reducing the error in Monte Carlo estimates. 
The main idea is to learn a control variate from the Monte Carlo draws, 
such that the distribution of the learnt function approximates that of the original problem. 
While doing so, it is crucial to avoid bias and overfitting, as this may worsen the final result.
We rewrite the solution to a Monte Carlo problem~$f(x)$ in terms a learnt function~$g(x)$
\begin{equation}\label{eq:StackMCInt}
\fHat =\int{f(x)p(x)\D x}
= \int{\alpha g(x)p(x)\D x} + \int{\left[f(x)-\alpha g(x)\right] p(x)}
= \alpha \gHat  + \int{\left[f(x)-\alpha g(x)\right] p(x)},
\end{equation}
where~$\alpha$ is some constant and $\gHat :=  \int{g(x)p(x)\D x}$. 
If~$g$ is a reasonable fit to~$f$, 
the difference $f - \alpha g$ for some appropriately chosen~$\alpha$ will have lower variance 
than the original problem. 
Over the observations $(x^{(i)})_{i=1,\ldots,N}$, the estimate
\begin{equation}\label{eq:methods:f_stack}
\fTilde := \alpha\gHat + \frac{1}{N}\sum_{i=1}^N \left[f\left(x^{(i)}\right) - \alpha g\left(x^{(i)}\right)\right]
\end{equation}
has variance $\sigma^2_{\fTilde} = \sigma^2_{f} + \alpha^2 \sigma^2_{g} - 2\alpha \sigma_{f,g}$
which is minimised as soon as 
\begin{equation}\label{eq:methods:alpha}
\alpha = \frac{\sigma_{f}}{\sigma_{g}}\rho_{f,g},
\end{equation}
which implies
$\sigma^2_{\fTilde} = (1-\rho_{f,g}^2)\sigma^2_{f}$,
and the condition $\rho_{f,g}\neq 0$ is enough to guarantee variance reduction. 
Intuitively, if~$g$ is a good fit then~$\rho$ should be high and the estimate~$\gHat$ is trusted, 
otherwise it is ignored. 
It remains to compute the function~$g$ and to estimate~$\alpha$, 
both of which are achieved by fitting the samples of~$f$.

%%%%%%%%%%%%%%%%%%%%%%%%%%%%%%%%%%%%%%%%%
\subsection{Model choice for the control variate}
In order to achieve significant reduction while controlling the computational cost, 
it is essential to choose a fitting function~$g$ for which the integral $\gHat$ in~\eqref{eq:methods:f_stack} 
is available in closed form, or at least is costless to compute. 
A convenient choice is a simple polynomial of order~$L$:
\begin{equation}
g(x) = \sum_{n=0}^{L}\ch_n x^n,
\end{equation}
where the coefficients $\chbf = (\ch_n)_{i=0, \ldots, L}$ 
is defined as the least-square minimiser
$$
\chbf := \argmin_{\cbf} \left\| f(x) - \sum_{n=0}^L c_n x^n \right\|^2.
$$
\begin{remark}\label{rem:Multig}
\begin{enumerate}[(i)]\ 
\item If the problem is multidimensional (in the case of basket options), 
with $\xbf = (x_i)_{i=1,\ldots,m}$, we may then consider a multivariate polynomial
$g(\xbf) = \sum_{|\alb|\leq L} c_{\alb}\xbf^{\alb}$,
using multi-index notations, where $\alb = (\alpha_1, \ldots, \alpha_m) \in \NN^m$.
\item Since Call option payoffs are discontinuous, 
it may be convenient to choose piecewise-polynomial functions, 
with zero value in part of the domain, for example, 
for some truncation plane~$\ab$,
\begin{equation}
g(\xbf) = \left\{
\begin{array}{ll}
0, & \text{if } \displaystyle \ab^T \xbf + \ab_0 < 0\\
 \displaystyle \sum_{|\alb|\leq L} c_{\alb}\xbf^{\alb}, & \text{if }\ab^T \xbf + \ab_0 \geq 0.
\end{array}\right.
\label{eq:methods:pl_g}
\end{equation}
\end{enumerate}
\end{remark}

%%%%%%%%%%%%%%%%%%%%%%%%%%%%
\subsection{Fitting the control variate}
Tracey et al.~\cite{Tracey} proposed using K-folds cross-validation to achieve an unbiased function estimate. This technique involves dividing the data into several non-overlapping training sets (or folds), 
and using the left out points to test the fit. 
We split the~$N$ samples of the variable~$\xbf$ into~$K$ folds of equal size 
to create the sets $\xbf^{(k),\text{train}}$ and $\xbf^{(k),\text{test}}$, 
for each $k=1,\ldots, K$. 
Each individual data sample is included in a test set once, and in a training set $K-1$ times. 
A different function $g_k(\xbf)$ is estimated using the~$N-N_k$ samples in each training set, 
and tested against the $N_k$ left-out samples from the $k$-th fold.
We thus obtain an estimate of the StackMC solution for each set from~\eqref{eq:methods:f_stack},
\begin{equation}\label{eq:methods:smc_k}
\fTilde_k := \alpha\gHat_k + \frac{1}{N_k}\sum_{i=1}^{N_k} 
\left[f\left(x_i^{(k),\text{test}}\right) - \alpha g_k\left(x_i^{(k),\text{test}}\right)\right],
\qquad\text{for }k=1,\ldots, K.
\end{equation}
And the final `stacked' solution is the average over the folds
\begin{equation}\label{eq:SMC}
\fTilde_{\text{SMC}} := \frac{1}{K} \sum_{k=1}^K \fTilde_k.
\end{equation}

%%%%%%%%%%%%%%%%%%%%%%%%%%%%%%%%%%%%%%%%%%%
\subsection{Estimating the control variate parameter} \label{sec:methods:alpha}
The parameter~$\alpha$ is estimated from the out-of-sample points 
using~\eqref{eq:methods:alpha} with the classical unbiased empirical estimates
$\widehat{\sigma}_{f,g}$, $\widehat{\sigma}^2_{f}$ and~$\widehat{\sigma}^2_{g}$ defined as
\begin{equation*}
\begin{array}{rlrl}
\widehat{\mu}_g & := \displaystyle \frac{1}{N}\sum_{k=1}^K \sum_{i=1}^{N_k} g_k\left(x_i^{(k),\text{test}}\right), & 
\widehat{\mu}_f & := \displaystyle \frac{1}{N}\sum_{i=1}^N f(x_i),\\
\widehat{\sigma}^2_{g}  & := \displaystyle \frac{1}{N-1}\sum_{k=1}^K \sum_{i=1}^{N_k} \left[g_k\left(x_i^{(k),\text{test}}\right)-\mu_g\right]^2,& 
\widehat{\sigma}^2_{f} & := \displaystyle \frac{1}{N-1} \sum_{i=1}^N (f(x_i) - \mu_f)^2, \\
\widehat{\sigma}_{f,g} & := \displaystyle \frac{1}{N-1} \sum_{k=1}^K\sum_{i=1}^{N_k}\left[f\left(x_i^{(k),\text{test}}\right) - \mu_f\right]\left[g\left(x_i^{(k),\text{test}}\right) - \mu_g\right].
\end{array}
\end{equation*}

 %%%%%%%%%%%%%%%%%%%%%%%%%%%%%%%%%%%%%%%%%
\subsection{Integrating the control variate}
We consider the computation of the analytical integral $\gHat$ of the control variate function in the case where~$g$ is a polynomial as in Remark~\ref{rem:Multig}(i):
$$
\gHat(\xbf) = \int\sum_{|\alb|\leq L} c_{\alb}\xbf^{\alb}p(\xbf)\D \xbf
 = \sum_{|\alb|\leq L} c_{\alb}\int\xbf^{\alb}p(\xbf)\D \xbf
 = \sum_{|\alb|\leq L} c_{\alb}\EE{\xbf^{\alb}}.
$$
In this case, the problem reduces to the computation  of moments.
If the random variable is drawn from a one-dimensional zero-mean Normal distribution with variance~$\sigma^2$, 
as in the standard Black-Scholes model, we may use the property:
\begin{equation}
\EE{X^n} = \left\{
\begin{array}{ll}
0, & \text{if }n\text{ is odd},\\
\displaystyle \frac{n! \sigma^n}{ 2 ^ {n / 2} (n/2)!}, & \text{if }n\text{ is even},
\end{array}
\right.
\end{equation}
In the multidimensional (centered) Gaussian case with variance-covariance matrix~$\Sigma$ 
(as will be useful later for Asian options), 
the moments can be derived from the moment generating function 
$\EE{\E^{\uu \Xb}} = \exp\left(\frac{1}{2}\uu^T\Sigma \uu\right)$.
In the case where the control variate has a piecewise linear form,
\begin{equation}
g(\xbf) = \left\{
\begin{array}{ll}
\displaystyle c_0 + \sum_j c_j x_j, & \text{if } \xbf \in \Omega_{+},\\
0, & \text{otherwise},
\end{array}\right.
\end{equation}
where $\Omega_{+}:= \left\{\xbf: \ab^T \xbf + a_0 \geq 0 \right\}$,
so that the computation of~$\gHat$ boils down to computing the zeroth and first moments of the truncated Gaussian distribution
$$
\int_{\Omega_+}{n(\xbf)\D \xbf}
\qquad\text{and}\qquad
\int_{\Omega_+}{x_j n(\xbf)\D \xbf},
\qquad\text{for each }j = 1,\ldots, M,
$$
where $n(\cdot)$ denotes the multivariate Gaussian density. 

\begin{remark}\label{rem:LinearCaseComput}
The integral above on a truncated domain can in fact be computed in closed form when the integrand
(without the Gaussian density) is linear, as here.
Following~\cite{Sharples}, since a linear combination of a multivariate Gaussian is Gaussian, we can write
$$
\int_{\Omega_+}n(\xbf)\D \xbf
 = \int_{-a_0}^{\infty} \frac{1}{\|\ab\|\sqrt{2\pi}}\exp\left\{-\frac{x^2}{2\|\ab\|^2}\right\}\D x
 = 1 - \Nn\left(-\frac{a_{0}}{\|\ab\|}\right).
$$
For the first moment, we can use~\cite[Theorem 5]{Sharples} to write
$$
\int_{\Omega_+}x_j n(\xbf)\D \xbf
 = \frac{a_j}{\left\|\ab\right\|\sqrt{2\pi}}\exp\left(-\frac{a_{0}^2}{2\|\ab\|^2}\right),
 \qquad\text{for any }j=1,\ldots, M.
$$
\end{remark}

%%%%%%%%%%%%%%%%%%%%%%%%%%%%%%%%%%%%%%%%%%%%
\section{Pricing options in the Black-Scholes model}
We now test the Stacked Monte Carlo method presented above on the pricing of options\footnote{All tests were conducted on a desktop with 2 Intel Xeon 2.00 GHz processors and 16 GB RAM, 
running Windows 7 Enterprise. The code was written in \texttt{Python~3.6}, using \texttt{Numpy~1.15} and \texttt{Scikit-learn~0.20}. For comparison purposes, all calculations were computed on a single thread.}.
We use the \texttt{random} method from \texttt{NumPy}, 
which employs a Mersenne-Twister generator, to generate all Gaussian samples. 
All numerical results reported here are obtained by running the Monte Carlo solver ten times 
and by taking the average of the parameter of interest (solution, confidence interval, relative improvement) over the runs. 
This avoids any bias with respect to the seed--determined by the system time from the pseudo-random number generator--especially for low numbers of paths. 

\subsection{Stacked Monte Carlo for European options}\label{sec:results_european}
We first consider the price of a European Call option with payoff $(S_T-K)_+$,
for some strike~$K$, when the underlying stock price~$S$ follows the Black-Scholes model
\begin{equation}\label{eq:BS}
\frac{\D S_t}{S_t} = r \D t + \sigma \D W_t,
\end{equation}
starting from $S_0>0$ for some one-dimensional Brownian motion~$W$.
To apply the StackMC algorithm, we note that
$\log(S_T)$ is a Gaussian random variable, and hence the price of the Call option is given by
$$
\E^{-rT}\EE{S_T - K}_+ = \E^{-rT}\int_{\RR}\left(S_0\E^{\left(r-\frac{\sigma^2}{2}\right)T + \sigma\sqrt{T}x} - K\right)_{+}n(x)\D x,
$$
where~$n(\cdot)$ denotes the Gaussian density, which is exactly of the form~\eqref{eq:StackMCInt}.
We start by fitting a polynomial function $g(\cdot)$ (of order $L=4$), 
and follow the StackMC methodology above with 
$K=2$ folds and $N=10^5$ Gaussian samples.
We then estimate~$\alpha$ as in Section~\ref{sec:methods:alpha}.
With the parameters:
$$
(S_0, K, r, \sigma, T) = (100, 100, 5\%, 20\%, 1),
$$
for which the exact price is $C_{\BS} = 10.4506$, the results are shown in Table~\ref{tab:OptionPrices}.

 % Table generated by Excel2LaTeX from sheet 'FirstResult'
\begin{table}[h]
\begin{tabular}{rrr|rrr|r|rr}
 	\multicolumn{ 3}{c|}{MC} & \multicolumn{ 3}{c|}{Stacked Monte Carlo} & \multicolumn{ 1}{c|} {Total} & \multicolumn{ 2}{c} {Improvement} \\
 	\hline
 	Price &         $\text{CI}_\text{MC}$ & Time &      Price &         $\text{CI}_\text{SMC}$ & Time & Time &   Absolute &   Ratio \\
 	\hline
 	10.4395 &     0.0913 &   0.57 &    10.4507 &     0.0061 &       0.30 &       0.87 &      0.0851 &     14.90 \\
 	\hline
 	\\
\end{tabular}
\caption{StackMC vs standard Monte Carlo for a European Call option}
\label{tab:OptionPrices}
\end{table}
Here, CI denotes the half-width of the confidence interval defined as 
$$
\left[\mu - u_{l/2}\frac{\sigma_N}{\sqrt{N}},\ \mu + u_{l/2} \frac{\sigma_N}{\sqrt{N}} \right],
$$
where~$\mu$ is the mean, $\sigma_N$ the sample standard deviation, and $u_{l/2}$ the Gaussian quantile ($u_{l/2} = 1.96$ for $l=95\%$). 
The absolute improvement is defined as $\text{CI}_\text{MC}-\text{CI}_\text{SMC}$, 
and the improvement ratio is $\text{CI}_\text{MC}/\text{CI}_\text{SMC}$.
All indicated times are in seconds.
To standardise the reporting of run times and render these independent of PC performance, 
we consider Table~\ref{tab:OptionPrices} as time unit, namely the time taken to price a European Call option with~$10^5$ Monte Carlo draws ($0.57$ second). 
We expect run times to scale (approximately) linearly with the number of simulations. 
Results below are reported in these units unless otherwise specified. 
We also quote an equivalent Monte Carlo time, defined as an estimate of the time 
it would have taken to achieve the confidence interval of StackMC using MC alone. 
This is the MC time multiplied by the square of the improvement ratio, 
and must be compared to the total runtime, i.e. the time taken to perform both MC and StackMC. 
Under such measure, the previous results read
\begin{equation*}
\begin{array}{cccc}
\text{MC} &    \text{StackMC} &  \text{Total} & \text{Equivalent} \\
1 &    0.52504 &     1.525 &     228.08\\
\end{array}
\end{equation*}

On average, the StackMC procedure achieved a nearly 15-fold improvement in the size of the confidence interval with respect to simple MC, at the cost of approximately $0.3$ seconds runtime. Considering that the Monte Carlo solution converges at a rate of~$\mathcal{O}\left(N^{-1/2}\right)$, to achieve a similar variance reduction using Monte Carlo alone, 
we must increase the number of simulations by a the improvement ratio squared (about~$228$).
Given that runtime scales linearly with the number of simulations, 
this would have taken significantly longer than the additional time taken by the stacking procedure, which is only about half of the MC runtime.
The code we are using here is not fully optimised, and significant speed improvements 
can be made for both the MC and StackMC implementations by exploiting parallelisation, 
minimising data loops, and increasing algorithm efficiency. 
Some speed-up gains may be greater for Monte Carlo, given the high potential for parallelisation~\cite{Joshi}.

%%%%%%%%%%%%%%%%%%%%%%%%%%%%%%%%%%%%%%%%%
\subsection{Dependency on model parameters}
\begin{figure}[h!]
	\subfigure[Size of the confidence interval]{\includegraphics[scale=0.4]{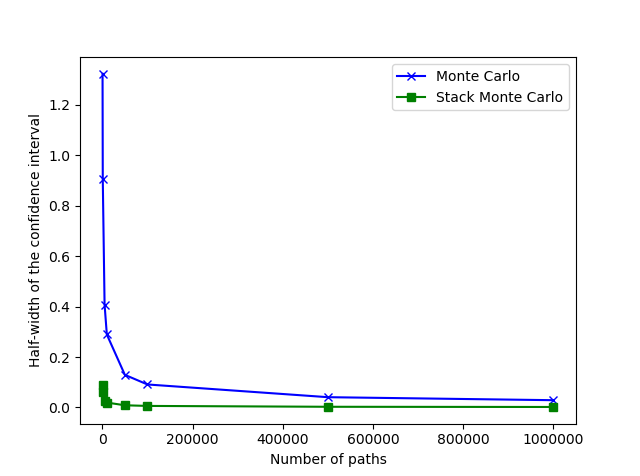}}
	\subfigure[Absolute confidence interval reduction]{\includegraphics[scale=0.4]{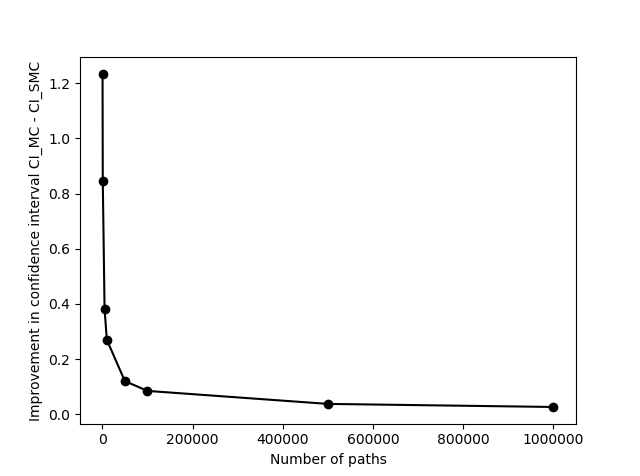}}
	\caption{Variance reduction for a European Call option, between MC and StackMC.}
	\label{fig:results:mcvsmc_nsteps}
\end{figure}

\begin{figure}[h!]
	\subfigure[Confidence interval reduction ratio]{\includegraphics[scale=0.4]{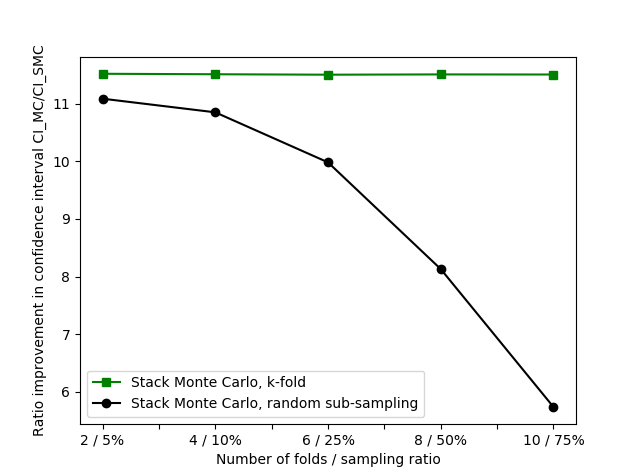}}
	\subfigure[Computation time of the stacking procedure]{\includegraphics[scale=0.4]{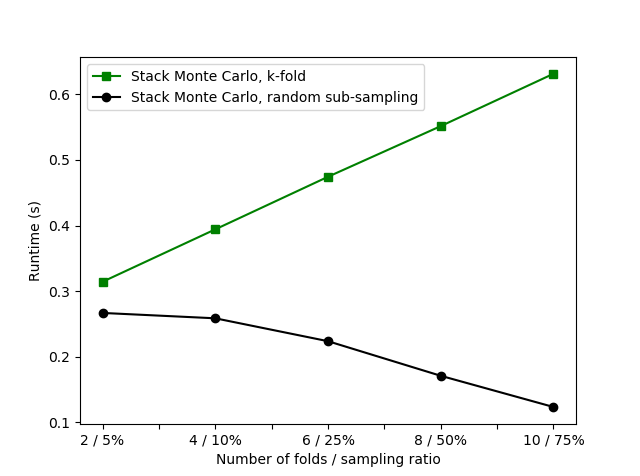}}
	\caption{Variance reduction by cross-validation between $K$-folds and random sub-sampling.}
	\label{fig:results:EURCrossVal}
\end{figure}

We now investigate how the number of paths (Table~\ref{tab:NbPaths} and Figure~\ref{fig:results:mcvsmc_nsteps}), cross-validation technique, 
and fitting model affect the result of Section~\ref{sec:results_european}, holding all else equal.
For the cross-validation method, we increase the number of folds used in the K-folds method, 
and compare the results with those obtained via simple random sub-sampling (without folds). 
We vary the proportion of data in the training sample, and use the remainder to estimate the control variate parameter~$\alpha$ (Figure~\ref{fig:results:EURCrossVal}). 

\begin{table}[h!]
% Table generated by Excel2LaTeX from sheet 'NumSamples'
\begin{tabular}{|r|rr|rr|rr|rr|}	
	\hline
 & \multicolumn{2}{|c|}{MC} & \multicolumn{2}{c|}{StackMC} & \multicolumn{2}{c|}{Improvement} & \multicolumn{2}{c|}{Time} \\
	\hline
Paths &  CI &  Time &    CI &    Runtime &   Absolute &      Ratio &      Total & Equivalent \\
\hline
$10^3$  &  0.9059 &   0.01  &     0.0601 &   0.01 &     0.08 &      15.18 &       0.02 &       2.27 \\
$10^4$  &  0.2895 &   0.10  &     0.0194 &    0.06 &     0.27 &      14.92 &       0.16 &      22.57 \\
$10^5$  &  0.0913 &   1.00  &     0.0061 &    0.53 &     0.09 &      14.90 &       1.53 &     222.08 \\
$10^6$  &  0.0288 &   10.15 &     0.0019 &   5.38 &    0.03 &      14.96 &      15.53 &    2271.72 \\
\hline
\multicolumn{9}{c}{}
\end{tabular}
\caption{Effect of the number of simulations on the variance reduction.}
\label{tab:NbPaths}
\end{table}

The Monte Carlo variance is reduced as expected at a rate of $\Oo\left(N^{-1/2}\right)$, and the StackMC variance is consistently lower (Figure~\ref{fig:results:mcvsmc_nsteps}). 
The improvement brought by the stacking procedure decreases as the number of simulations increases, and is much more stable.
In Figure~\ref{fig:results:EURCrossVal}, 
we observe a very modest improvement when increasing the number of folds. 
The computational expense increased with the number of folds, 
due in most part to the time spent fitting all the functions~$g_k$. 
The random sub-sampling performs worse than K-folds in all cases, and performance depends of the balance between data used to estimate the model parameters (in-sample data), and to calculate~$\alpha$ (out-of-sample data). Runtime is lower than that of K-folds, and decreases as the proportion of training data increased, 
largely due to the shorter time spent estimating~$\alpha$ on a smaller test set.

%%%%%%%%%%%%%%%%%%%%%%%%%%%%%%%%%%
\subsubsection{Choice of fit model} \label{sec:results_fit_model}
\begin{figure}[!htbp]
	\subfigure[Size of the confidence interval]{\includegraphics[scale=0.4]{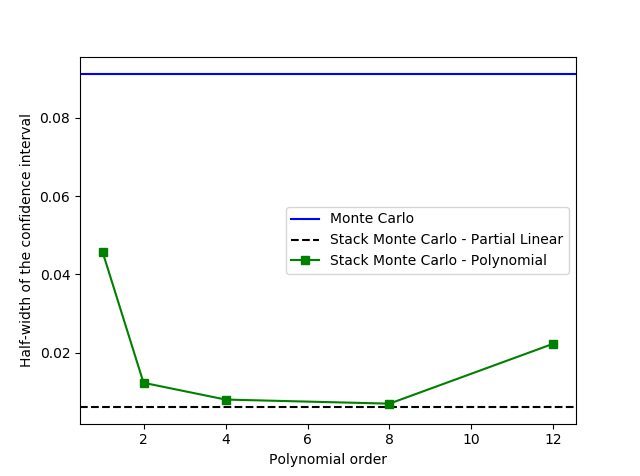}}
	\subfigure[Confidence interval reduction ratio]{\includegraphics[scale=0.4]{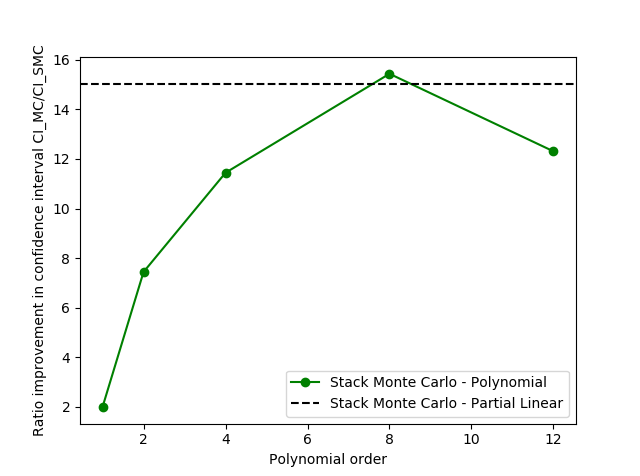}}
	\caption{Variance reduction between MC and StackMC by fit model for a European Call option.}
	\label{fig:results:mcvsmc_poly}
\end{figure}

To investigate the effect of the fit model, we perform the stacking procedure using polynomial functions~$g(\cdot)$ 
of increasing order, and compare these results with those obtained for a piecewise linear fitting function (Figure~\ref{fig:results:mcvsmc_poly}). 
Rather than performing a non-linear fit to determine the inflection point, 
the piecewise function is obtained by filtering the zero-valued training data, 
and by estimating a linear fit to the remaining points. 
Values to the left of the intercept with the horizontal axis are then set to zero.
We observe a large improvement between orders~$1$ and~$2$, 
and small improvements thereafter up to order~$8$, after which the results worsen 
(Figure~\ref{fig:results:mcvsmc_poly}). 
On average, the performance of the piecewise linear fitter is comparable to that of the polynomial functions (Figure~\ref{fig:results:mcvsmc_poly}).
It is worth noting that our method for finding the piecewise linear function described above 
was chosen for computational speed. 
For comparison purposes, we also applied a non-linear method (using \texttt{Scikit-learn}) to fit a generic piecewise-linear function with one inflection point, 
and found the gains in variance reduction to be almost identical, 
albeit with a much higher computational cost. 

%%%%%%%%%%%%%%%%%%%%%%%%%%%%%%%%%%
%%%%%%%%%%%%%%%%%%%%%%%%%%%%%%%%%%
\subsection{Pricing Asian options}

\begin{figure}[!htbp]
\subfigure[Polynomial of order 2.]{
\includegraphics[scale=0.32]{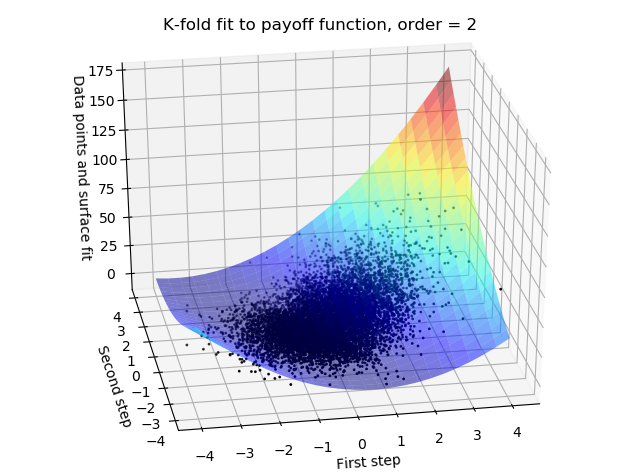}}
\subfigure[Polynomial of order 4.]{
\includegraphics[scale=0.32]{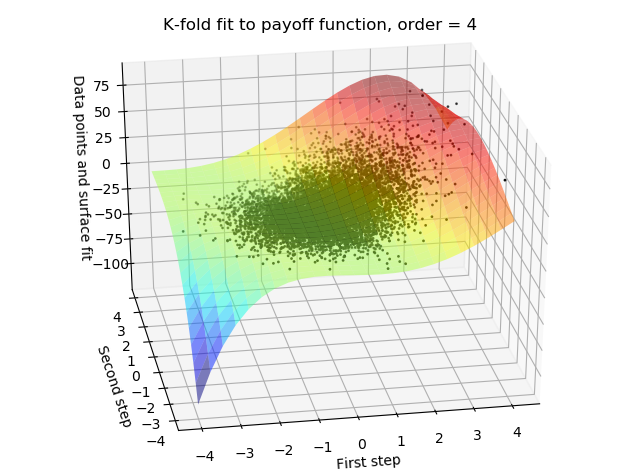} }
\subfigure[Piecewise linear]{
\includegraphics[scale=0.32]{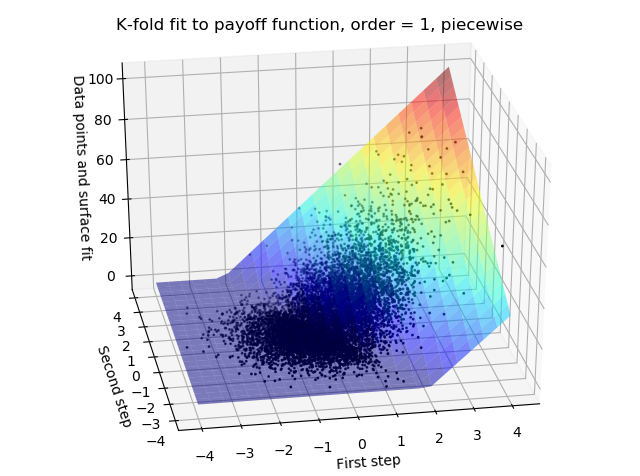}}
\caption{StackMC variance reduction on an Asian option with $M=2$.}
\label{fig:results:asians_toy}
\end{figure}

%%%%%%%%%%%%%%%%%%%%%%%%%%%%%%%%%%
We are now interested in testing the Stacked Monte Carlo procedure on some path-dependent option,
and we consider the case of an arithmetic Asian option, whose payoff is given by
$$
\Phi_T := \left(\frac{1}{M}\sum_{i=1}^{M}S_{t_i} - K\right)_+,
$$
for some strike~$K$ and some monitoring dates $0<t_1<\ldots<t_{M}$.
Under no-arbitrage arguments, the price of the Asian option reads, at time zero,
$\E^{-rT}\EE{\Phi_T}$.
In the Black-Scholes model~\eqref{eq:BS}, by independence of the Gaussian increments, we can write
(with $t_0=0$)
$$
\sum_{i=1}^{M}S_{t_i}
 = S_{t_0}\sum_{i=0}^{M-1}\prod_{j=0}^{i}\frac{S_{t_{j+1}}}{S_{t_j}}
   = S_0\sum_{i=0}^{M-1}\prod_{j=0}^{i}\exp\left\{\left(r-\frac{\sigma^2}{2}\right)(t_{j+1}-t_{j}) + \sigma\sqrt{t_{j+1}-t_{j}}X_j\right\}
   =: f(X_1,\ldots, X_M),
$$
where $(X_1,\ldots, X_M)$ is a centered Gaussian vector with identity covariance matrix.
The fitting problem is thus of dimension~$M$, where the variables~$X_1, \ldots, X_M$ 
correspond to the Brownian increments. 
With the same parameters as in Section~\ref{sec:results_european}, 
Figure~\ref{fig:results:asians_toy} shows the results of the StackMC 
procedure using polynomial surfaces of degrees~$2$ and~$4$, 
and a piecewise linear surface as in Section~\ref{sec:results_fit_model}. 

Tables~\ref{tab:AsianComparre2}, \ref{tab:AsianComparre50} and~\ref{tab:AsianComparre100} illustrate the method with
respectively~$M=2$, $M=50$ and~$M=100$, in order to to illustrate the effect of the dimensionality. 
In this example, the advantage of using a piecewise linear fitter is apparent: 
not only does this scheme deliver a greater improvement in the variance, 
but the computational effort required is such that it enables scaling of the problem to high dimensions. 
This property is essential to the application of the StackMC methodology to Asian options, 
for which each path is defined by an $M$-dimensional vector of Brownian increments.
The runtime and memory requirements increase greatly when the number of time steps increases, 
and the performance of the second-order polynomial fit is consistently worse than that of the piecewise linear function.

\begin{table}[!htbp]
	% Table generated by Excel2LaTeX from sheet 'AsianToy'
	\begin{tabular}{rrrrr|r|rr|}		
		\cline{6-8}
		\multicolumn{ 1}{c}{} &            &            &            &            & Improvement & \multicolumn{ 2}{c|}{Time} \\
		\cline{2-8}
		\multicolumn{ 1}{c|}{} &  Fit model &      Price &         CI &    Runtime &      Ratio &      Total & Equivalent \\
		\hline
		\multicolumn{ 1}{|r|}{MC} &            &     8.1024 &     0.0698 &      21.36 &            &            &            \\
		\hline
		\multicolumn{ 1}{|c|}{} &  Polynomial 2  &     8.1108 &     0.0097 &       1.23 &     7.18 &      22.59 &    1100.43 \\
		
		\multicolumn{ 1}{|r|}{StackMC} &  Polynomial 4  &     8.1112 &     0.0064 &       1.54 &    10.97 &      22.90 &    2570.85 \\
		
		\multicolumn{ 1}{|c|}{} &  Piecewise Linear  &     8.1117 &     0.0042 &       1.27 &    16.71 &      22.63 &    5963.16 \\	
		\hline
		\\
	\end{tabular}  
	\caption{Comparison of MC and StackMC for the Asian option with $M=2$.}
		\label{tab:AsianComparre2}
\end{table}

\begin{table}[!htbp]
	% Table generated by Excel2LaTeX from sheet 'AsianToy'
	\begin{tabular}{rrrrr|r|rr|}		
		\cline{6-8}
		&            &            &            &            & Improvement & \multicolumn{ 2}{c|}{Time} \\
		\cline{2-8}
		\multicolumn{ 1}{c|}{\multirow{1}{*}{}} &  Fit model &      Price &         CI &    Runtime &      Ratio &      Total & Equivalent \\
		\hline
		\multicolumn{ 1}{|r|}{MC} &            &     5.8532 &     0.0502 &      22.24 &            &            &            \\
		\hline
		\multicolumn{ 1}{|r|}{\multirow{2}{*}{StackMC}} &  Polynomial 2  &     5.8582 &     0.0073 &     102.64 &       6.87 &     124.88 &    1049.08 \\
		
	\multicolumn{ 1}{|c|}{}	&  Piecewise Linear  &     5.8567 &     0.0502 &       3.10 &      19.88 &      25.34 &    8785.44 \\	
	\hline
	\\
	\end{tabular}  
		\caption{Comparison of MC and StackMC for Asian option pricing with $M=50$
		(the memory requirements for the 4th-order surface fit is prohibitive, and hence omitted).}
		\label{tab:AsianComparre50}
\end{table}

\begin{table}[!htbp]
% Table generated by Excel2LaTeX from sheet 'AsianToy'
\begin{tabular}{rrrrr|r|rr|}
		\cline{6-8}	
	\multicolumn{ 1}{c}{} &            &            &            &            & Improvement & \multicolumn{ 2}{c|}{Time} \\
	\cline{2-8}
	\multicolumn{ 1}{c|}{} &  Fit model &      Price &         CI &    Runtime &      Ratio &      Total & Equivalent \\
	\hline
	\multicolumn{ 1}{|r|}{MC}  &            &     5.8209 &     0.0499 &      23.10 &            &            &            \\
	\hline
	\multicolumn{ 1}{|r|}{\multirow{2}{*}{StackMC}} &  Polynomial 2  &     5.8100 &     0.0076 &    1149.01 &       6.60 &    1172.12 &    1005.32 \\
	
	\multicolumn{ 1}{|c|}{} &  Piecewise Linear  &     5.8102 &     0.0025 &       5.59 &      19.84 &      28.69 &    9092.11 \\
		\hline
	\\
	\end{tabular}  
		\caption{Comparison of MC and StackMC for Asian option pricing with $M=100$.}
		\label{tab:AsianComparre100}
\end{table}

\vspace{1cm}

%%%%%%%%%%%%%%%%%%%%%%%%%%%%%%%%%%%%%%%%
We add one final numerical example (Table~\ref{tab:AsianComparre365}) for the Asian case, 
in line with market considerations, 
namely with a piecewise linear model and daily time intervals ($M=365$ and $T=1$ year). 
\begin{table}[!htbp]
% Table generated by Excel2LaTeX from sheet 'AsianReal'
\begin{tabular}{|r|rrr|rrr|rr|rr|}
	\hline
	&    \multicolumn{ 3}{|c}{Monte Carlo} & \multicolumn{ 3}{|c}{Stacked Monte Carlo} & \multicolumn{ 2}{|c}{Improvement} & \multicolumn{ 2}{|c|}{Time} \\
	\hline
	Paths &      Price &         CI &       Time &      Price &         CI &       Time &        Abs &      Ratio &      Total & Equivalent \\
	\hline
	1E4  &     5.7867 &     0.1570 &       2.97 &     5.7759 &     0.0084 &       2.10 &     0.0149 &      18.77 &       5.07 &    1045.88 \\
	
	1E5 &     5.7727 &     0.0496 &      29.56 &     5.7762 &     0.0025 &      18.12 &     0.0471 &      19.84 &      47.68 &   11635.73 \\
	
	1E6 &     5.7755 &     0.0157 &     299.46 &     5.7759 &     0.0008 & 199.37 &     0.0149 &      19.92 &     498.83 &  118783.86 \\
	\hline
	\multicolumn{11}{c}{}
\end{tabular}  
		\caption{Comparison of MC and StackMC for Asian option pricing with $M=365$.}
		\label{tab:AsianComparre365}
\end{table}
For $N=10^5$ Monte Carlo simulations, 
the additional time spent on the stacking procedure is about~$18$ in our time units 
(equivalent to about ten seconds), 
which yields a nearly 20-fold improvement in the variance.

%%%%%%%%%%%%%%%%%%%%%%%%%%%%%%%%%%%%%%
%%%%%%%%%%%%%%%%%%%%%%%%%%%%%%%%%%%%%%
\section{Stacked Monte Carlo for stochastic volatility models}
\label{sec:results_stochvol}
We now adapt the Stacked Monte Carlo method to European and Asian options in local stochastic volatility models of the form
\begin{equation}\label{eq:SDEStochVol}
\left\{
\begin{array}{rll}
\displaystyle \frac{\D S_t}{S_t} & = \displaystyle r \D t + \sigma_{\mathrm{loc}}(S_t)\xi(t, V_t)\D W_t, & S_0>0,\\
\D V_t & = \displaystyle b(t, V_t)\D t + a(t, V_t)\D B_t, & V_0>0,
\end{array}
\right.
\end{equation}
where~$r$ denotes the risk-free interest rate, and~$W, B$ are two Brownian motions with correlation~$\varrho\in [-1, 1]$.
The coefficients~$b(\cdot)$, $a(\cdot)$, $\sigma_{\mathrm{loc}}(\cdot)$ and $\xi(\cdot)$ are left undefined,
and are such that a unique solution to the system exists.
Sets of sufficient conditions are classical, and can be found in~\cite[Chapter 5]{Karatzas} for example.

%%%%%%%%%%%%%%%%%%%%%%%%%%%%%%%%%%%%%%
\subsection{Methodology}
We start by discretising the system~\eqref{eq:SDEStochVol} following some Euler scheme.
For the purpose of our methodology, any convergent schemes suffices,
and an overview of such discretisations can be found in~\cite{Kloeden}.
In particular, we start by drawing two matrices of standard Gaussian samples~$\Wf\in\Mm_{N,M}$
and~$\Bf\in\Mm_{N,M}$
(corresponding to the standardised Brownian increments of the stock price and the variance respectively), 
such that the correlation between~$\Wf_{ij}$ and ~$\Bf_{ij}$
is equal to~$\rho$, for any $i=1,\ldots, N$ and $j=1,\ldots, M$,
where~$N$ and~$M$ respectively denote the number of paths and the number of time steps.
For each path, we can then compute the price of the option under consideration;
we shall denote by $\Pif = (\Pi_1, \ldots, \Pi_N)$ the corresponding vector of prices.
Dividing the sample into $K$ folds of sizes $N_1, \ldots, N_K$, 
we denote by $\Pif^k \in \RR^{N_k}$ the part of the vector~$\Pif$ corresponding to the $k$-th fold and by $\CPi^k \in \RR^{N-N_k}$ the vector~$\Pif$ without~$\Pif^k$
($k=1, \ldots, K$), and similarly for~$\Wf^k \in\Mm_{N_k,M}$ and $\breve{\Wf}^k\in\Mm_{N-N_k,M}$ (column-wise).
Up to reordering, we assume that the elements of~$\Pif$ and the rows of~$\Wf$ are ordered, 
so that the first~$N_1$ of them correspond to the first fold, and so on.
For any $k=1, \ldots, K$, a predictor~$\overline{\Pif}^k$ of~$\Pif^k$ is then obtained as
$$
\overline{\Pif}^k\left(\Wf^k\right) := \Pp\left(\Wf^k\right),
$$
where~$\Pp$ denotes a polynomial of any order, the coefficients of which are calibrated
through the fit of~$\breve{\Pif}^k$ on the cloud of points~$\breve{\Wf}^k$.
Note again that we do not use the $k$-th fold for the regression, but only for prediction 
(the in-sample data).
This fit is essentially a supervised machine learning problem that can be solved 
either by least-square regressions or other techniques such as neural networks or random forests~\cite{Hastie}.
In order to set up the control variate, for each fold~$k$, we compute the integral
\begin{equation}\label{eq:ControlVarHeston}
\widehat{\Pif}^k := \int \overline{\Pif}^k\left(\Wf^k\right) n(\Wf^k)\D\Wf^k
 = \int \Pp\left(\Wf^k\right)n(\Wf^k)\D\Wf^k,
\end{equation}
and the version of the control variate~\eqref{eq:methods:smc_k} in the present context reads
$$
\widetilde{\Pif}^k := \frac{1}{N_k}\sum_{j=1}^{N_k}\Pif^k_j
 + \alpha\left(\widehat{\Pif}^k - \frac{1}{N_k}\sum_{j=1}^{N_k}\overline{\Pif}^k\left(\Wf^k_j\right)\right)
$$
With our notations, we have $\Pif^k_j = \Pif_{N_1 + \cdots+N_{k-1} + j}$;
the final stacked estimator corresponding to~\eqref{eq:SMC} is then
$$
\widetilde{\Pif}_{\text{SMC}} := \frac{1}{K} \sum_{k=1}^K \widetilde{\Pif}^k.
$$
On the numerical side, the computation of~\eqref{eq:ControlVarHeston}
may not be that straightforward.
However, in the spirit of Remark~\ref{rem:LinearCaseComput},
we restrict the integration domain to
$\left\{\Wf^k: \Pp\left(\Wf^k\right)\geq 0\right\}$, which is natural as the payoff should remain positive
for usual type of options such as Calls and Puts 
(other constraints can be considered should one be interested in other types of payoff functions).
The linearity of~$\Pp$ as well as this truncation domain thus yield the closed-form expressions 
from Remark~\ref{rem:LinearCaseComput} for the integral~\eqref{eq:ControlVarHeston}.

\begin{remark}
In~\eqref{eq:SDEStochVol}, we could replace the Brownian motion~$B$ by
a more general continuous Gaussian process~$G$ such as a fractional Brownian motion
or a Gaussian Volterra process, in the spirit of the recent rough volatility wave~\cite{RoughVol1, RoughVol2, RoughVol3, RoughVol4, RoughVol5}.
Since any continuous Gaussian Volterra process~$G$ 
has a representation of the form $G_t = \int_{0}^{t}K(s,t)\D B_s$, for some Brownian motion~$B$
defined on the same filtration as~$G$ and some kernel~$K(\cdot)$, 
then the knowledge of the increments~$\Bf$ of~$B$ provides the increments of~$G$, 
and the Stacked Monte Carlo methodology above still applies.
\end{remark}

%%%%%%%%%%%%%%%%%%%%%%%%%%%%%%%%%%%%%%
\subsection{Application to the Heston model}
In order to motivate our results numerically, 
we specialise~\eqref{eq:SDEStochVol} to the Heston~\cite{Heston} model, 
under which the stock price satisfies the system
\begin{equation}\label{eq:Heston}
\left\{
\begin{array}{rll}
\displaystyle \frac{\D S_t}{S_t} & = \displaystyle r \D t + \sqrt{V_t}\D W_t, & S_0>0,\\
\D V_t & = \displaystyle \kappa(\theta-V_t)\D t + \xi\sqrt{V_t}\D B_t, & V_0>0,
\end{array}
\right.
\end{equation}
for some parameters $ \kappa, \theta, \xi>0$, where~$W$ and~$B$
are two Brownian motions with correlation~$\varrho\in [-1, 1]$.
Several Euler schemes exist in the literature for the Heston model, 
in particular keeping track of the necessary positivity of the variance process, and we refer the interested reader to~\cite{Alfonsi, Alfonsi1, Alfonsi2} for an overview. 
We consider the following set of parameters:
$$
(S, K, r, T, V_0, \kappa, \theta, \xi, \varrho) = 
(100, 100, 3.19, 1, 1.02, 6.21, 1.9, 0.61, -0.7),
$$
for which the reference price for of the European Call option, as reported in~\cite{Broadie}, is equal to $6.8061$. 
Table~\ref{tab:HestonStackEuro} and Table~\ref{tab:HestonStackAsian} show the results of the procedure with $M=365$ time steps.
The results are not as clear as before, and only lead to a modest improvement in the variance. 

\begin{table}[h!]
% Table generated by Excel2LaTeX from sheet 'StochVol'
\begin{tabular}{|r|rrr|rrr|rr|rr|}	
\hline
	&    \multicolumn{ 3}{|c|}{Monte Carlo} & \multicolumn{ 3}{|c}{Stacked Monte Carlo} & \multicolumn{ 2}{|c}{Improvement} & \multicolumn{ 2}{|c|}{Time} \\
	\hline
	Nb Simulations &      Price &         CI &       Time &      Price &         CI &       Time &        Abs &      Ratio &      Total & Equivalent \\
	\hline
1E4  &     6.7930 &     0.1454 &      58.22 &     6.8158 &     0.0905 &       2.13 &     0.0549 &       1.61 &      60.35 &     150.20 \\
5E4 &     6.8197 &     0.0652 &     290.58 &     6.8113 &     0.0388 &       8.70 &     0.0263 &       1.68 &     299.29 &     818.58 \\
1E6  &     6.8118 &     0.0460 &     600.01 &     6.8088 &     0.0272 &      18.33 &     0.0188 &       1.69 &     618.34 &    1712.07 \\
\hline
	\multicolumn{11}{c}{}
\\
	\end{tabular}
\caption{StackMC for a European option in the Heston model.}
\label{tab:HestonStackEuro}
\end{table}

\begin{table}[h!]
% Table generated by Excel2LaTeX from sheet 'StochVol'
% Table generated by Excel2LaTeX from sheet 'StochVol'
\begin{tabular}{|r|rrr|rrr|rr|rr|}	
	\hline
	&    \multicolumn{ 3}{c|}{Monte Carlo} & \multicolumn{ 3}{c|}{Stacked Monte Carlo} & \multicolumn{ 2}{c|}{Improvement} & \multicolumn{ 2}{c|}{Time} \\
	\hline
	Nb Simulations &      Price &         CI &       Time &      Price &         CI &       Time &        Abs &      Ratio &      Total & Equivalent \\
	\hline
	1E4  &     3.6115 &     0.0762 &      59.28 &     3.6150 &     0.0469 &       2.09 &     0.0293 &       1.62 &      61.37 &     156.46 \\
	
	5E4  &     3.6222 &     0.0342 &     304.00 &     3.6182 &     0.0200 &       7.93 &     0.0142 &       1.71 &     311.92 &     887.88 \\
	
	1E5  &     3.6159 &     0.0241 &     611.52 &     3.6173 &     0.0140 &      18.63 &     0.0101 &       1.72 &     630.14 &    1809.95 \\
	\hline
	\multicolumn{11}{c}{}
	\\
\end{tabular} 
\caption{StackMC for an Asian option in the Heston model.}
\label{tab:HestonStackAsian}
\end{table}

\begin{figure}[h!]
	\subfigure[Constant volatility.]{
	\includegraphics[scale=0.3]{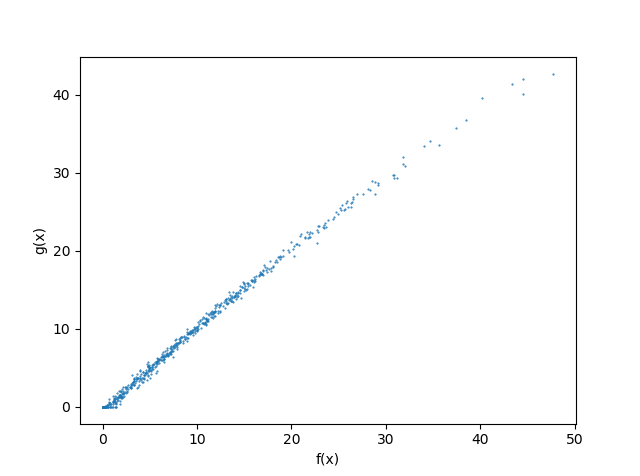}}
	\subfigure[Stochastic volatility.]{
	\includegraphics[scale=0.3]{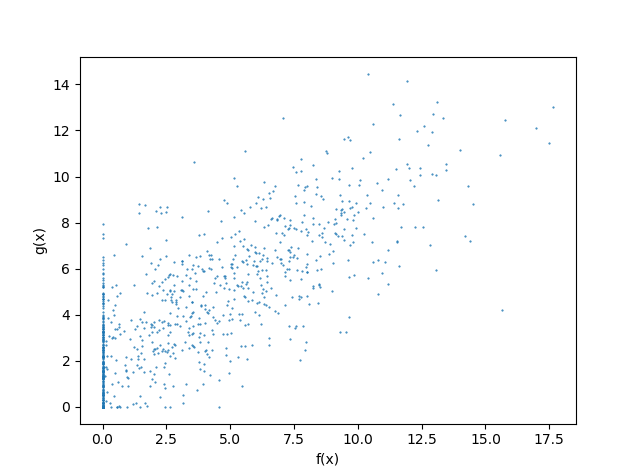} }\label{fig:results:stochvolb}
	\subfigure[Stochastic volatility.]{
	\includegraphics[scale=0.3]{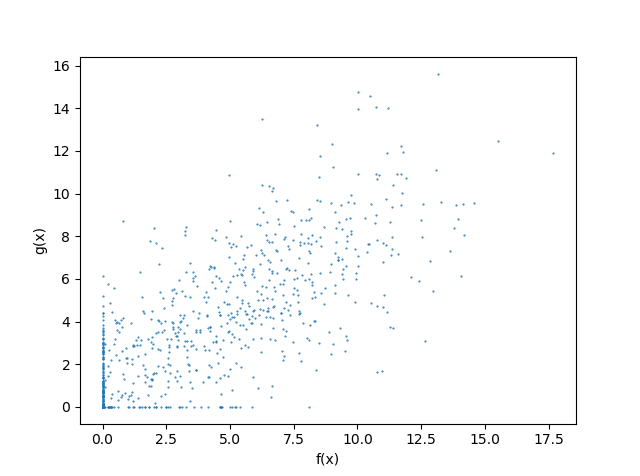}}
	\caption{Correlation between the data~$f$ and the control variate~$g$ for Asian options; 
	(a) and~(b) are obtained by filtering non-zero data and performing a linear fit, 
	while~(c) is determined by fitting a piecewise linear function to all the data using a non linear method.}
	\label{fig:results:stochvol}
\end{figure}

An example of the agreement between the payoff function and the control variate is displayed in Figure~\ref{fig:results:stochvol}. 
We note that the correlation is much higher in the case of constant volatility. 
Further, when a piecewise linear fit is achieved by filtering the non-zero data and fitting a linear function to the remainder, 
the fit appears biased toward smaller estimates (Figure~\ref{fig:results:stochvol}(b)). 
This is because in the presence of the variance-produced `noise', 
there are data points drawn for negative~$\xbf$-values with a (small) positive payoff, 
which are not being filtered, biasing the fit to the left. 
This effect could be mitigated by fitting a piecewise linear function to the full dataset 
using a non-linear method (Figure~\ref{fig:results:stochvol}(c)). 
Similarly, a larger variance reduction might be obtained by fitting a generic function, 
such as an $n^\text{th}$ order polynomial. 
However we are restricted by computational constraints to considering simple linear functions.

%%%%%%%%%%%%%%%%%%%%%%%%%%%%%%%%%%%%%%
\subsection{Comparison with other variance reduction methods}
We finally compare the results achieved using StackMC with those obtained with other commonly employed variance reduction methods. 
We concentrate on European and Asian Call options in the Black-Scholes model~\eqref{eq:BS}
as in Section~\ref{sec:results_european}
In all cases we draw~$100,000$ simulations, apply K-fold cross validation with $K=2$, 
and choose a piecewise linear control variate function (fitted with a linear method to the positive data).

%%%%%%%%%%%%%%%%%%%%%%%%%%%%%%%%%%%%%%
\subsubsection{Antithetic updates}
We compare StackMC to a standard Monte Carlo method with antithetic updates~\cite{Glasserman}.
With the same values as in Section~\ref{sec:results_european}, 
From the results in Table~\ref{tab:StacKMCAnti},
we see that StackMC achieves much greater variance reduction, for both European and Asian options. 

\begin{table}[h!]
% Table generated by Excel2LaTeX from sheet 'VarReduction'
\begin{tabular}{r|rr|rr|rr|rr|}
	&   \multicolumn{ 4}{c|}{European option} & \multicolumn{4}{c|}{Asian option} \\
	\cline{2-9}
		&   \multicolumn{ 2}{c|}{Result} & \multicolumn{ 2}{c|}{Improvement} 	&  \multicolumn{ 2}{|c|}{Result} & \multicolumn{ 2}{|c|}{Improvement}\\
	\cline{2-9}
	&      Price &         CI &        Abs &      Ratio 	&      Price &         CI &        Abs &      Ratio\\
	\hline
	MC  &    10.450 &     0.0914 &            &             &     &     &            &             \\
	\hline
	Antithetic MC  &    10.452 &     0.0646 &     0.0268 &       1.41 &     5.7803 &     0.0351 &     0.0145 &       1.41\\
	{}
	StackMC &     10.452 &     0.0061 &     0.0852 &      14.90 &     5.7763 &     0.0025 &     0.0471 &      19.75\\
	\hline
	\multicolumn{9}{c}{}
	\\
\end{tabular} 
\caption{StackMC with antithetic updates.}
\label{tab:StacKMCAnti}
\end{table}

%%%%%%%%%%%%%%%%%%%%%%%%%%%%%%%%%%%%%%%%%%%%
\subsubsection{Geometric mean control variate}
For an arithmetic Asian option in the Black-Scholes models, it is common practice to use 
the geometric version as a control variate~\cite{Joshi, Kemna}, 
so that the original pricing problem is replaced with
\begin{equation}\label{eq:results:geom_cv}
\widehat{C}_{\text{A}} := \left(C_{\text{A}} - C_{\text{G}}\right) + \widehat{C}_{\text{G}},
\end{equation}
with~$C_{\text{A}}$ and~$C_{\text{G}}$ the arithmetic and Geometric Call option prices.
The difference $ (C_{\text{A}} - C_{\text{G}})$ is computed using the Monte Carlo simulations.
Note that $C_{\text{G}}$ is available in closed form, and is equal to~\cite{Kemna}
$$
C_{\text{G}} = \E^{-rT}\left[S_0\E^{aT}\Nn(d_+) - K\Nn(d_-)\right],
$$
with
$$
d_{\pm} := \frac{\log(S_0/K) + \left(a\pm \frac{\sigma^2}{3}\right)T}{\sigma\sqrt{T}/3}
\qquad\text{and}\qquad
a := \frac{1}{2}\left(r-\frac{\sigma^2}{2}\right).
$$
The comparison between StackMC and the geometric mean as control variate for an Asian option with parameters as in Section~\ref{sec:results_european} and~$M=365$ is shown in 
Table~\ref{tab:CampreStackGeom}.

\begin{table}[h!]
% Table generated by Excel2LaTeX from sheet 'VarReduction'
\begin{tabular}{r|rr|rr|}	
	& \multicolumn{ 2}{c|}{Result} & \multicolumn{ 2}{c|}{Improvement} \\
	\cline{2-5}
	&      Price &         CI &        Abs &      Ratio \\
	\hline
	MC  &     5.7811 &     0.0050 &            &            \\
	\hline
	CV MC  &     5.7795 &     0.0023 &     0.0473 &      21.71 \\
	
	StackMC &     5.7758 &     0.0025 &     0.0470 &      19.82 \\	
	\hline
	\multicolumn{5}{}\\
\end{tabular}  
\caption{Comparison of StackMC and MC with Geometric control variate.}
\label{tab:CampreStackGeom}
\end{table}

The variance reduction achieved with the geometric mean is slightly larger than that of StackMC. However, we may also apply Stacked MC to the new control variate 
problem~\eqref{eq:results:geom_cv} and compound the improvements. 
The results of the StackMC applied to the modified problem~\eqref{eq:results:geom_cv} are summarised in Table~\ref{tab:GeomMeanStackMCControl}.
\begin{table}[h!]
 	% Table generated by Excel2LaTeX from sheet 'VarReduction'
 	\begin{tabular}{r|rr|rr|}	
 		& \multicolumn{ 2}{c|}{Result} & \multicolumn{ 2}{c|}{Improvement} \\
 		\cline{2-5}
 		&      Price &         CI &        Abs &      Ratio \\
 		\hline
 		\multicolumn{1}{|c|}{StackMC+CV}  &     5.7796 &     0.0011 &     0.0485 &      46.12 \\
 		\hline
 		\multicolumn{5}{}
 		\\
 	\end{tabular}  
 	\caption{StackMC on the modified problem.}
 	\label{tab:GeomMeanStackMCControl}
 \end{table}

\begin{figure}[h!]
	\includegraphics[scale=0.5]{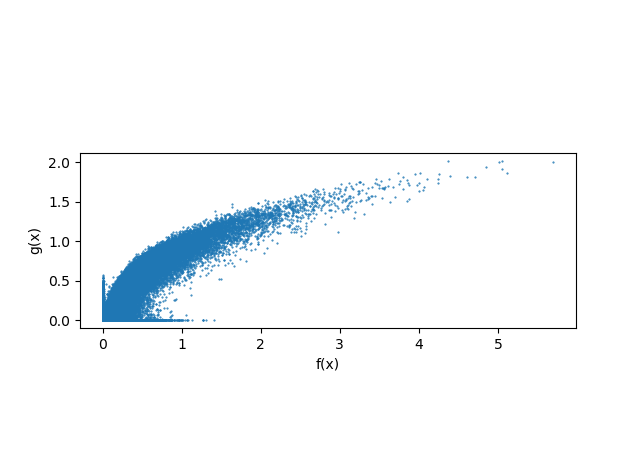}
	\caption{Correlation between the data and the fit function for the modified Asian option pricing problem using the Geometric mean as control variate.}
	\label{fig:results:varreduction}
\end{figure}

The advantage of the stacking method is that whilst using the geometric mean is only successful in the particular case of the arithmetic Asian option, StackMC has general validity and can be applied (in theory) to any problem. 
As such, it applies to the control variate modified problem~\eqref{eq:results:geom_cv}, 
which delivers an additional compounded variance reduction (Figure~\ref{fig:results:varreduction}). 

%%%%%%%%%%%%%%%%%%%%%%%%%%%%%%%%%%%%%%%%%%%

%%%%%%%%%%%%%%%%%%%%%%%%%%%%%%%%%%%%%%%%%%%%%%

\end{document}